\documentclass[11pt,a4paper]{article}
\usepackage{jheppub}
\pdfoutput=1
\usepackage{epsfig}
\usepackage{latexsym}
\usepackage{amsfonts}
\usepackage{amsmath}
\usepackage{amsthm}
\usepackage{amssymb}
\usepackage{amsbsy}
\usepackage{multirow}
\usepackage{slashed}
\usepackage{physics}

\newcommand{\CC}{\mathbb{C}} 
\newcommand{\RR}{\mathbb{R}} 
\newcommand{\NN}{\mathbb{N}} 

\def\calf         {{\cal F}}

\def\caln         {{\cal N}}
\def\calo         {{\cal O}}

\def\be{\begin{equation}}
\def\ee{\end{equation}}
\def\bea{\begin{eqnarray}}
\def\eea{\end{eqnarray}}

\def\a{\alpha}
\def\b{\beta}

\def\d{\delta}
\def\e{\epsilon}

\def\l{\lambda}

\def\f{\phi}

\def\m{\mu}

\def\o{\omega}

\def\p{\pi}
\def\r{\rho}

\def\s{\sigma}

\def\F{\Phi}
\def\pa{\partial}

\def\to{\rightarrow}
\def\nonu{\nonumber \\{}}
\def\half{{1 \over 2}}

\def\mh{\mathfrak{h}}
\def\ma{\mathfrak{a}}

\title{On  matter coupled to the higher spin square}
\author{Joris Raeymaekers}

\affiliation{Institute of Physics of the ASCR, \\
Na Slovance 2, 182 21 Prague 8, Czech Republic.}

\emailAdd{joris@fzu.cz}

\abstract{Gaberdiel and Gopakumar recently proposed that the tensionless limit of string theory on $AdS_3 \times S^3 \times T^4$
takes the form of a higher spin theory with a gauge algebra that is referred to as  the higher spin square. In this note, we
formulate the linearized Vasiliev-type equations which describe a matter field coupled to the higher spin square. We study the particle spectrum of this field and  show that it accounts for the entire  untwisted sector of the dual symmetric orbifold CFT, thereby confirming a conjecture by  Gaberdiel and Gopakumar. In doing so, we  pinpoint the group-theoretic data which determine  the spectrum of a matter field  coupled to  a general higher spin algebra, which we illustrate   by revisiting the theory based on the $hs[1/2]$ algebra.} \preprint{????}

\begin{document}
\maketitle

\section{Introduction}
The idea that string theory may possess a phase of enhanced higher spin gauge symmetry is a longstanding one \cite{Vasiliev:1995dn}. String field theory would   most likely take its simplest form when formulated around the vacuum of maximal unbroken  symmetry,  where it would be maximally constrained by an underlying gauge principle. Phases with massive higher spin fields, such as the standard formulation around the Minkowski background, would then arise as Higgsed phases upon expanding around a vacuum which breaks the higher spin symmetry. The work of Vasiliev  (see e.g. \cite{Vasiliev:1999ba}, \cite{Didenko:2014dwa} for  reviews)  has shown that  interacting higher spin gauge theories allow for an  anti-de Sitter vacuum, and  it is therefore natural to look for the most symmetric phase of string theory in an AdS compactification.

Though conceptually appealing, these ideas didn't receive a concrete realization until the recent work of Gaberdiel and Gopakumar \cite{Gaberdiel:2014cha},\cite{Gaberdiel:2015mra},\cite{Gaberdiel:2015wpo} (see also the
related works \cite{Gaberdiel:2013vva},\cite{Gaberdiel:2014yla},\cite{Gaberdiel:2015uca}). They proposed the tensionless limit of string theory on $AdS_3 \times S^3 \times T^4$ as a candidate for the phase with maximal unbroken symmetry.  It has a  dual description in terms of  a two-dimensional superconformal field theory: the $N$-fold symmetric product orbifold  of the $\caln = 4$ SCFT on the torus $T^4$, denoted by $Sym^N (  T^4 )$, in the limit of large  $N$.  From analyzing the  dual theory they were able to isolate the symmetry algebra which plays the role of the gauge algebra in the bulk. This algebra, which contains exponentially more generators than the standard higher spin  algebras, is nevertheless fully determined by the action of  standard  `horizontal' and 'vertical' higher spin algebras and therefore goes under the name of the higher spin square ($hss$) \cite{Gaberdiel:2015mra}. We will review its definition in section \ref{Sechss} below. Furthermore,  the spectrum of the untwisted sector of $Sym^N (  T^4 )$  has a simple content  in terms of $hss$ representations.

In this note, we will take a first step towards reformulating tensionless string theory in the bulk  as a higher spin gauge theory with $hss$ symmetry. In particular, we will show that the entire spectrum of the untwisted sector arises, on the bulk side, from a single matter field coupled to the $hss$ gauge fields, thereby   confirming a conjecture made in \cite{Gaberdiel:2015wpo}. Along the way, we will single out the group-theoretic data which determine  the spectrum of single-particle states described by  Vasiliev-like matter equations  in $AdS_3$  \cite{Vasiliev:1992gr} for a general higher spin algebra. We will illustrate  this by revisiting the Vasiliev theory based on the higher spin algebra $hs[1/2 ]$ from this point of view.

Let us first collect  some of the results of \cite{Gaberdiel:2015wpo}. It was shown there  that the NS-NS sector  partition function of
the untwisted sector of $Sym^N (  T^4 )$ can, at large $N$, be written a way which is suggestive of a dual bulk interpretation:
 \bea
 Z_U (q, \bar q, y, \bar y) &\equiv& \tr_U q^{L_0} \bar q^{\tilde L_0}  y^{2 J^3_0} \bar y^{2 \tilde J^3_0} \\
&=&   Z^{gauge}_{hss}(q, \bar q, y, \bar y)  Z_{hss}^{matter}(q, \bar q, y, \bar y) .\label{ZU}
 \eea
 Here, $Z_{hss} $  is the mod-squared of the vacuum character of the chiral algebra of the dual CFT
 \be
Z^{gauge}_{hss}(q, \bar q, y, \bar y) =  |Z_{vac}(q,y)|^2.\label{Zgauge}
 \ee
 In analogy with the Vasiliev higher spin theories \cite{Gaberdiel:2010ar}, this contribution is expected to come   from  boundary excitations of   gauge fields taking values in the higher spin square Lie algebra $hss$.
 The second factor  in (\ref{ZU}) is suggestive of a contribution from  matter in the bulk. It takes the `multi-particle' form\footnote{The minus signs in this expression  account for statistics, as we will explain in more detail in section \ref{Sechss} below.}
 \be
 Z_{hss}^{matter} (q, \bar q, y, \bar y)  = \exp \sum_{n=1}^\infty {Z_{hss}^{1-part} ( q^n, \bar q^n, (-1)^{n+1} y^n, (-1)^{n+1}\bar y^n ) \over n}\label{Zmatter}
 \ee
 and appears to describe
 multi-particle excitations of a theory whose single particle spectrum is given by
\be
Z_{hss}^{1-part} (q, \bar q, y, \bar y) 
= |\chi_{min}(q,y)|^2
 \label{Z1part}
\ee
Here, $\chi_{min}$ is the character of the so-called minimal representation \cite{Gaberdiel:2014cha} of the higher spin square. More explicitly, it is given by
\be
 \chi_{min}(q,y) =Z_ {T^4}(q,y) - 1\label{chimin}
 \ee
where $Z_{T^4}$ is the chiral part of the partition function of the SCFT on $T^4$, i.e. the
partition function of 4 real chiral bosons and 4 Mayorana-Weyl fermions:
\be
Z_{T^4}(q,y) = \prod_{n=1}^\infty {( 1+ y q^{n-\half})^2( 1+ y^{-1} q^{n-\half})^2\over (1-q^n)^4}.\label{ZT4}
\ee
Our goal in this note will be   to show how (\ref{Z1part}) arises from single-particle excitations of matter fields in the bulk.

\section{Free Vasiliev system and single particle states}

In this section we will review  the linearized equations describing  matter coupled to massless   higher spin fields in 2+1 dimensions.  These equations were originally \cite{Vasiliev:1992gr} written down  for the higher spin theories with higher spin algebra $hs[\l]$  (and their supersymmetric extensions), but we will not yet specify the higher spin algebra here, emphasizing instead the algebraic ingredients necessary to write down a consistent set of equations. This will pave the way for analyzing matter fields coupled to the higher spin square in section \ref{Sechss}.
\subsection{Massless higher spin fields}
Massless higher spin fields in 2+1 dimensions are described by a gauge theory based on two copies\footnote{See however \cite{Arvanitakis:2015sgs} for an example of a higher spin theory based on a Lie algebra which is simple, rather than a product of two isomorphic copies.} of a higher spin Lie algebra $\mh$, with $\mh$-valued gauge potentials  $A, \tilde A$. We will assume $\mh$ to contain an
$sl(2,\RR)$ subalgebra, which we will single out as the subsector which describes   Einstein gravity in $AdS_3$. In general, the spin content of the theory is determined by the decomposition of the adjoint representation of $\mh$ into representations of this $sl(2,\RR)$ subalgebra.
The equations of motion state that $A, \tilde A$ are flat:
 \begin{align}
d  A -  A \wedge  A &=0, &
d  \tilde A - \tilde A \wedge \tilde  A &=0.\label{F0}
\end{align}
Hence the gauge sector doesn't contain any local propagating degrees of freedom, although  a careful treatment of the boundary conditions at infinity \cite{Henneaux:2010xg},\cite{Campoleoni:2010zq} generically  uncovers the existence of boundary excitations.

The equations of motion are formally invariant under  finite higher spin gauge transformations of the form
\begin{align}
A&\to hA h^{-1} + dh h^{-1}, &
\tilde A&\to \tilde h \tilde A \tilde h^{-1} + d\tilde h \tilde h^{-1}\label{Agt}
\end{align}
where $h, \tilde h$  belong to $H$, the set of formal exponentials of elements of $\mh$. We will not address here the question for which elements of $\mh$ the exponential is well-defined, nor whether $H$ can be given the structure of of a Lie group. In the case where $\mh$ is the higher spin algebra  $hs[\l ] $, these issues were addressed in \cite{Monnier:2014tfa}.
\subsection{Linearized matter equations}
Next we review the linearized equations describing matter in a higher spin background specified by $A, \tilde A$.
In most known examples, the higher spin Lie algebra $\mh$ can be embedded in an associative algebra $\ma$, such that the Lie bracket in $\mh$  arises from the commutator in $\ma$.
If this is the case, we can write down linearized  equations\footnote{In the
 original work \cite{Vasiliev:1992gr}, see also \cite{Ammon:2011ua}, the equations were written  more succinctly by introducing an extra Grassmann element $\f$, satisfying $\f^2=1$, and the associated projection operators $P_\pm = {1 \pm \f\over 2}$.
 Combining the fields as  $W= A P_+ + \tilde A P_-, B = C P_+  + \tilde C P_-$, the equations (\ref{F0},\ref{Ceqgen}) reduce to
 \begin{align} d W - W\wedge W &=0, &\qquad
d B - W B + B \p ( W )&=  0\nonumber \end{align}
where the operation $\p$ sends $\f \to - \f$.}  for two scalar matter fields $C, \tilde C$  taking values in the associative algebra $\ma$:
\begin{align}
d  C -  A  C + C \tilde  A &=0 &
d  \tilde C -  \tilde A \tilde  C + \tilde  C   A &=0 .\label{Ceqgen}
\end{align}
Here the fields  are multiplied using the associative product in $\ma$.
Note that the consistency of (\ref{Ceqgen}) is guaranteed by (\ref{F0}).
These equations are usually written in an equivalent `star-product' form,
in which the $\ma$-valued fields  are replaced by  $c$-number functions (or `symbols') by using a specific operator ordering prescription,
and the effect of the operator product is captured by
 a suitably defined star-product. We will however keep working with $\ma$-valued fields  in this note.
The equations (\ref{Ceqgen}) are higher spin gauge invariant with the fields transforming as
 \begin{align}
 C &\to h C \tilde h^{-1},&
 \tilde C &\to \tilde h \tilde C h^{-1}\label{gaugetransf}
\end{align}
When $\mh$ is one of the standard $hs[\l ]$ higher spin algebras and $\ma$ is the underlying `lone-star product' associative algebra \cite{Pope:1989sr}, the equations  (\ref{Ceqgen}),
when expanded around the AdS background,  describe the unfolded form of the Klein-Gordon equation for a  scalar field of mass squared $\l^2-1 $. We will review this in some detail  for the case of $\l = \half$ in section \ref{Sechshalf} below.

One might think that it is consistent to truncate the theory to keep only the $C$ or $\tilde C$ matter field. We will see however in the next section
that $C$ describes only negative frequency modes, while $\tilde C$ describes positive frequency ones. In order to have a sensible phase
space lending itself to quantization, we should therefore keep both $C$ and $\tilde C$ even at the linearized level.

\subsection{The  AdS background}
The advantage of the unfolded form of the equations is that we can easily write down the general solution to  (\ref{Ceqgen}) in any background \cite{Vasiliev:1992gr}.
 Indeed, writing the flat connections $A, \tilde A$ locally in a pure gauge form,
\begin{align}
A &= dg g^{-1},& \tilde A &= d\tilde g \tilde g^{-1}
 \end{align}
we can transform to a gauge where $A = \tilde A=0$. In this gauge the solutions for the matter fields are simply constant elements $C^0, \tilde C^0 \in \ma$. Introducing a basis $\{ e_a \}_a$ for $\ma$
and gauge-transforming back,  we can write the general solution to (\ref{Ceqgen}) as a linear combination of the solutions
\begin{align}
C_a &= g e_a \tilde g^{-1},&
\tilde C_a &= \tilde  g e_a  g^{-1}.\label{gensol}
\end{align}

Let us now specialize to the global AdS background, for which $g$  is an exponential of elements in the $sl(2,\RR)$ subalgebra of $\mh$
with commutation relations $[L_m,L_n] = (m-n) L_{m+n}$ for $m,n = 0, \pm1$.
Separating out the dependence on the $AdS$ radial coordinate $\r$, the group elements can be written conveniently as \cite{Perlmutter:2012ds}
\begin{align}
g &= R(\r ) e^{iL_0 x_+},&
\tilde g =& \tilde R(\r ) e^{-iL_0 x_-},\label{ggt}
\end{align}
where $x_\pm = t\pm \f$ and
\be R(\r ) =  e^{-\r L_0}M^{-1}, \qquad \tilde R(\r ) =  e^{\r L_0}M^{-1}, \qquad M= e^{- {i \p\over 4}(L_1 - L_{-1})}.\ee
The presence  of the constant element $M$ is for later convenience: it implements a change of basis which diagonalizes the  element $\half (L_1 + L_{-1})$, in the sense that
\be
M \cdot \half (L_1 + L_{-1}) \cdot M^{-1} = - i L_0.
\ee
Plugging this decomposition into (\ref{gensol}), the mode solutions for $C$ and $\tilde C$ around $AdS$ are given by
 \begin{align}
C_a &= R e^{iL_0 x_+} e_a e^{iL_0 x_-} \tilde R^{-1},&
\tilde C_a =& \tilde R e^{-iL_0 x_-} e_a e^{-iL_0 x_+}  R^{-1}.\label{gensolAdS}
\end{align}

  \subsection{Higher spin representations and single-particle spectrum}
  We would also like to determine how the solutions (\ref{gensolAdS}) transform under the global higher spin symmetries of the $AdS$ background. The unfolded formulation makes symmetries manifest in general (see e.g. \cite{Vasiliev:2012vf}), and we will now argue that the symmetry properties of the solutions (\ref{gensolAdS}) follow easily once a certain group-theoretic fact about the algebra $\ma$ is known.

  First let us discuss the global symmetries of the $AdS$ background and their action on the matter fields \cite{Vasiliev:1992gr},\cite{Perlmutter:2012ds}. As before it is convenient to transform to the gauge  $A= \tilde A=0$ where the matter fields are constant. Global symmetries are  gauge transformations leaving the background $A= \tilde A=0$ invariant; from (\ref{Agt}) these  are generated by constant elements
  $h_0, \tilde h_0 $ and hence the global symmetry consists of two copies of $H$, which we will denote as  $H \times \tilde H$. These  global symmetries act on  constant matter solutions $C^0 , \tilde C^0\in \ma$ by left-and right multiplication according to (\ref{gaugetransf}):
   \begin{align}
   C^0 & \to h_0 C^0 \tilde h_0^{-1}, & \tilde  C^0 & \to \tilde  h_0 \tilde C^0 h_0^{-1}, & h_0 \in H , \tilde h_0 \in \tilde H.\label{globalA0}
   \end{align}
    We can view the algebra $\ma$ as a representation space for  $H \times \tilde H$ with, say, $\tilde H$ acting from the left and $ H$ acting from the right.
    Therefore we expect to be able to decompose $\ma$ into irreducible $H \times \tilde H$ representations as follows
   \be
   \ma = \oplus_i (V_i, W_i).\label{decomp}
   \ee
   This decomposition  determines  how the space of matter solutions decomposes into irreducible representations of the global symmetry in the gauge $A = \tilde A=0$.

 These observations can then be simply gauge transformed to the `AdS gauge' where
 \be A = dg g^{-1}, \qquad \tilde A = d \tilde g \tilde g^{-1}\label{AdSgauge}\ee
 with $g, \tilde g$ given in (\ref{ggt}). The elements which implement the global $H \times \tilde H$ symmetry in this gauge are simply obtained from (\ref{globalA0}) by conjugation with $g$ and $\tilde g$:
   \begin{align}
   h &= g h_0 g^{-1},&  \tilde h &= \tilde g \tilde h_0 \tilde g^{-1}.
   \end{align}
   Therefore the decomposition (\ref{decomp}) also determines
 the quantum numbers carried by solutions in the AdS gauge (\ref{AdSgauge}).
   As a check on our analysis so far, let us work out the infinitesimal gauge parameters implementing the  $SL(2,\RR )\times \widetilde{SL(2,\RR )}$ subgroup of the global symmetry:
 \begin{align}
 \e_{L_0} &= b^{-1} {i\over 2}( L_1 +L_{-1}) b, &  \e_{L_{\pm1}} &= b^{-1} e^{\mp i x_+} \left(i L_0  \pm \half (L_1 - L_{-1}) \right) b \nonu
\tilde \e_{\tilde L_0} &= b {i\over 2}( L_1 +L_{-1}) b^{-1}  , & \tilde \e_{\tilde L_{\pm1}} &= b e^{\pm i x_-} \left(i L_0  \pm \half (L_1 - L_{-1}) \right) b^{-1}
  \end{align}
  where $b = e^{\r L_0}$.
These agree, modulo differences in conventions,  with the expressions derived  in \cite{Gaberdiel:2011wb} in the  AdS gauge (\ref{AdSgauge}).

 We have argued that the decomposition (\ref{decomp}) determines how the solution space to (\ref{Ceqgen}) decomposes into representations of the global symmetry. In free field theory,
  the space of solutions generally decomposes into positive and negative frequency modes, and only one of these subspaces (usually chosen to be the positive frequency subspace) can be given the structure of a Hilbert space
  representing the single particle states of the theory. For example, for a complex scalar field, the Hilbert space inner product is constructed from the conserved $U(1)$ current, and a higher spin invariant conserved current also exists for the linearized
  Vasiliev system based on the $hs[\l ]$ algebra \cite{Prokushkin:1999xq} (see also\cite{Vasiliev:2002fs}).
  It would be interesting to generalize this construction to the higher spin square theory,
  but in   what follows  we will simply assume the usual correspondence between positive  frequency modes and single particle states.

From (\ref{gensolAdS}), we see that, if the representations entering in the decomposition (\ref{decomp}) are unitary (as will be the case in our examples), so that $L_0$ has positive eigenvalues, the positive
frequency modes are contained in the field $\tilde C$, while  $C$  describes the  negative frequency ones. Therefore only  the modes of $\tilde C$ represent single particle states, leading to a partition function
of the form
\bea
Z^{1-part}(q,\bar q) &=& \tr_\ma q^{L_0} \bar q^{\tilde L_0}\\
&=& \sum_i \chi_{V_i} (q) \chi_{W_i} (\bar q),\label{1partspectr}
\eea
where $L_0$ and $\tilde L_0$ denote quantum numbers under right- and left action respectively. In the last line we have used (\ref{decomp}) to write the result in terms of characters
$\chi_{V_i}$  of the representations $V_i$.
A similar formula holds if we refine the partition function with chemical potentials.

To summarize, we have argued that the decomposition (\ref{decomp}) contains the group-theoretic data which determine the matter spectrum for generic higher spin algebras.
In the next two sections we will work out  (\ref{decomp}), first for the warm-up example of the theory based on the $hs[1/2 ]$ algebra, and subsequently for the case of interest where $\mh$ is the
higher spin square $hss$. Before doing so it may be instructive to work out (\ref{decomp}) for some finite-dimensional examples. Let $\ma$ the algebra
of real $5\times 5$ matrices with the standard matrix multiplication. Let $\mh$ be the subalgebra $sl(2,\RR)$ whose generators are embedded as $L_m
= L_m^{({\bf 2})} \oplus L_m^{({\bf 3})}$, with $L_m^{({\bf n})}$ the generators in the {\bf n}-dimensional representation. It's easy to see that $\ma$ decomposes into  $SL(2,\RR) \times \widetilde{SL(2,\RR)}$ representations as
\be
\ma = 
 ({\bf 2}, {\bf 2}) \oplus ({\bf 2}, {\bf 3}) \oplus({\bf 3}, {\bf 2}) \oplus({\bf 3}, {\bf 3}).
\ee
If, on the other hand, $\mh$ is $sl(2,\RR)$ embedded as $L_m
= L_m^{({\bf 5})}$, the decomposition is simply
\be
\ma = 
 ({\bf 5}, {\bf 5}).
\ee

\section{Matter coupled to {\em hs[1/2]}}\label{Sechshalf}
As a warmup to the higher spin square case, let us discuss the decomposition (\ref{decomp})
 in the  case where $\mh$ is the Lie algebra $hs[1/2 ]$ and $\ma$ is the underlying `lone-star algebra'  $\ma_{hs[1/2]}$ introduced in  \cite{Pope:1989sr}.
This example is instructive since, as will be the case for the higher spin square, $\ma_{hs[1/2]}$ has a realization in terms of undeformed harmonic oscillators. Furthermore, we will be able to give a clearer
 physical justification of one of the steps involved in this simplified setting.

We start from a single harmonic oscillator
\be
[a, a^\dagger] =1. \label{acomm}
\ee
Here, $a$ and $a^\dagger$  can be thought of as related to the  Vasiliev operators $y_1,y_2$ as $a = {y_1\over \sqrt{2 i}}, a^\dagger =  {y_2\over \sqrt{2 i}}$.
The standard basis of $\ma_{hs[1/2]}$ is  formed by Weyl-ordered monomials of even degree in the oscillators,
\be
V^s_m = 2^{1-s} \left[ (a^\dagger)^{s-m-1} a^{s+m-1} \right]_W, \qquad  s\geq 1, |m|<s.\label{weylbasis}
\ee
where the subscript stands for Weyl-ordering. In particular, $V^1_0$ is the identity operator, and the remaining generators $V^s_m$ with $s\geq 2$ generate, through their commutators, the Lie algebra $hs[1/2 ]$.
It was shown in \cite{Ammon:2011ua} that multiplying these basis elements and using the commutation relation (\ref{acomm}) to once again Weyl order the result reproduces the lone-star product of
$\ma_{hs[1/2]}$ defined in \cite{Pope:1989sr}.
The $s=2$ generators form an $sl(2, \RR) $ subalgebra, with  $L_0 = V^2_0,\ L_{\pm 1} = V^2_{\pm 1}$. One can check that the quadratic Casimir $C_2$ of this $sl(2, \RR) $ subalgebra takes the value
\be
C_2 = L_0^2 - \half \left(L_1 L_{-1} + L_{-1} L_1 \right)= -{3 \over 16}.\label{Casrel}
\ee
This observation allows one to alternatively  view $\ma_{hs[1/2]}$ as the quotient of the universal enveloping algebra $U(  sl(2, \RR) )$ by the ideal $C_2 + 3/16$.

The basis (\ref{weylbasis}) of Weyl-ordered monomials  is not very suitable for figuring out the desired decomposition of the form (\ref{decomp}). For this purpose, one would like to diagonalize the action of the Cartan generators $V^s_0$ both from the left and the right, while one easily checks that this is not the case in the basis  (\ref{weylbasis}). To remedy this we will instead expand operators in a Fock  basis\footnote{A similar basis change in the case of four-dimensional higher spin theories was advocated in \cite{Iazeolla:2008ix}.}, in terms of basis elements
\be
\ket{m}\bra{n} = {(a^\dagger)^m \over \sqrt{m!} } \ket{0}\bra{0} {a^n \over \sqrt{n!} } \equiv {1\over \sqrt{m! n!}} \left[(a^\dagger)^m a^n\right]_F.\label{Fockbasis}
\ee
In the last equality we have defined, for later convenience,  a `Fock-ordering'   acting on oscillator monomials by letting the creation and annihilation operators act on $\ket{0}  \bra{0} $ from the left and  the right respectively.
The relations between Weyl-ordered and Fock-ordered  monomials is  summarized by the following  formulae, whose origin is explained in Appendix \ref{App}:
\bea
\, \left[(a^\dagger)^m a^n\right]_W &=& \left\{ \begin{array}{lcr} 2^{-n}n! \left[   (a^\dagger )^{m-n} L^{m-n}_n (- 2 a a^\dagger ) e^{a a^\dagger}\right]_F &\  & {\rm for}\ m\geq n\\
 2^{-m}m! \left[  a^{n-m} L^{n-m}_m (- 2 a a^\dagger )e^{a a^\dagger}\right]_F&\ & {\rm for}\ n\geq m \end{array} \right.\label{WeyltoFock}\\
\, \left[(a^\dagger)^m a^n \right]_F &=& \left\{ \begin{array}{lcr}(-1)^n 2^{m-n+1}n! \left[   (a^\dagger )^{m-n} L^{m-n}_n (4 a a^\dagger ) e^{-2a a^\dagger}\right]_W &\  & {\rm for}\ m\geq n\\
(-1)^m 2^{n-m+1}m! \left[  a^{n-m} L^{n-m}_m (4 a a^\dagger )e^{-2 a a^\dagger}\right]_W&\ & {\rm for}\ n\geq m \end{array} \right. ,\label{FocktoWeyl}
\eea
where $L^k_n$ are the associated Laguerre polynomials.  Since the right-hand side of these
equations is  non-polynomial, one might question the validity of transforming  to the Fock basis, and we will provide a physical justification for this step below.

From these expressions it follows that even Weyl-ordered monomials are expressed in terms of even Fock-ordered monomials and vice versa.  Therefore
(\ref{WeyltoFock},\ref{FocktoWeyl}) allow us to represent the standard basis elements ${V^s_m}$ of $\ma_{hs[1/2]}$ in terms of the subset op  Fock basis elements
\be  \ket{m} \bra{n}
{\rm \ with \ } m+n {\rm \ even},\label{abasisfock} \ee
and vice versa. For example one finds, from (\ref{WeyltoFock}),
\bea
1 &=& V^1_0 = \sum_n \ket{n} \bra{n} \\
L_0 &=& V^2_0 = \half \sum_n \left( n + \half\right) \ket{n} \bra{n} \label{L0hshalf}\\
L_1 &=& V^2_1 = \half \sum_n \sqrt{(n+1)(n+2)} \ket{n} \bra{n+2} \label{L1hshalf}\\
L_{-1} &=& V^2_{-1} = \half \sum_n \sqrt{(n+1)(n+2)} \ket{n+2} \bra{n} .\label{sl2hshalf}
\eea

Using the Fock basis  of $\ma_{hs[1/2]}$ we can now now easily derive the decomposition of the form (\ref{decomp}). Decomposing the harmonic oscillator Fock space $\calf$ into
\be
\calf = \calf_{even} \oplus \calf_{odd},
\ee
 where $\calf_{even}$ ($\calf_{odd} $) are spanned by the basis states with even (odd)  excitation number $\ket{2m}$ ($\ket{2m+1}$) respectively, we have, due to (\ref{abasisfock}),
 \be
 \ma_{hs[1/2]} = \left( \calf_{even} \otimes \calf_{even}^* \right) \oplus \left( \calf_{odd} \otimes \calf_{odd}^* \right).
\ee
where $\,^*$ denotes the dual vector space.
The subspaces $\calf_{even} $ and $\calf_{odd}$ each form irreducible lowest weight  representations of $hs[1/2 ]$ whose lowest weight vectors are $\ket{0}$ and $\ket{1}$ respectively.
Indeed, from  (\ref{L1hshalf}) we see that they are annihilated by $L_1$, 
and that their  $L_0$ eigenvalues are ${1 \over 4}$ and ${3\over 4}$ respectively. Using the description of  $hs[1/2 ]$  as a quotient of the universal enveloping algebra of $sl(2,\RR) $ by the relation (\ref{Casrel}), one can show
  that $\ket{0}$ and $\ket{1}$  are also lowest weight vectors under the full $hs[1/2]$ algebra.  The corresponding irreducible representations
were denoted by $\f_-$ and $\f_+$ respectively in \cite{Gaberdiel:2011wb}. We also note that writing  the $hs[1/2 ]$ elements in the Fock basis as in (\ref{sl2hshalf}) and restricting their action to either $\calf_{even}$ or $\calf_{odd}$, we obtain the infinite-dimensional matrix representations of $hs[1/2 ]$  discussed  in \cite{Khesin:1994ey},\cite{Hijano:2013fja},\cite{Campoleoni:2013lma}.

It follows  that, for  $hs[1/2 ]$, the decomposition (\ref{decomp}) reads,
\be
\ma_{hs[1/2]} = \left( V_{\f_-} , V_{\f_-}^* \right) \oplus \left( V_{\f_+} , V_{\f_+}^* \right).\label{decomphshalf}
\ee
where $\,^*$ denotes the dual representation.
Applying (\ref{1partspectr}), the spectrum of single particle states described by the linearized Vasiliev equations is
\be
Z^{1-part}_{hs[1/2]} = |\chi_{\f_-} |^2 + |\chi_{\f_+} |^2.
\ee
This agrees with the conclusions reached in \cite{Vasiliev:1992gr} using  somewhat different methods.

Let us now discuss in more detail the explicit solutions which furnish the representations in the decomposition (\ref{decomp}). These will turn out to be
 physically sensible, which provides  additional justification for the initial step of expanding in the Fock  basis (\ref{Fockbasis}).
 From (\ref{gensolAdS}), the first term in (\ref{decomphshalf}) is furnished by  the positive frequency solutions of the form
\be
\tilde C_{2m,2n}^{AdS} = e^{-i\left( (m+ n + \half) t + (n-m)\f\right) }\tilde R \ket{2m}\bra{2n} R^{-1}\label{soleven}
\ee
while the second term is realized on the solutions
\be
\tilde C_{2m+1,2n+1}^{AdS} = e^{-i\left( (m+ n + {3\over 2}) t + (n-m)\f\right)} \tilde R \ket{2m+1}\bra{2n+1} R^{-1}.\label{solodd}
\ee
 The physical content of the matter fields $C, \tilde C$  is contained in the component  of the unit operator
  $V^1_0$ when expanded in the basis (\ref{weylbasis}), while the other components in this expansion are auxiliary fields \cite{Vasiliev:1992gr}.  The operation of extracting the  component of $V^1_0$ is referred to as taking the trace, and we will denote it by  $\tr$, though it should not be confused with the trace operation in the Hilbert space of the harmonic oscillator. The equations of motion (\ref{Ceqgen}) imply that the physical fields \be \F = \tr C, \qquad \tilde \F= \tr \tilde C\ee
  satisfy the Klein-Gordon equation with $m^2= - 3/4$ \cite{Vasiliev:1992gr}.
    One finds that the physical wavefunctions of the solutions (\ref{soleven}, \ref{solodd}) are, up to an inconsequential normalization factor, given by
  \bea
\tilde \F_{2m, 2n} &=& e^{-i \left( (n+{1 \over 4}) x_+ + (m+{1/4})x_-\right)} {\rm tr} \left( e^{-i\r  (L_1 + L_{-1})} \ket{2m}\bra{2n} \right)\nonu
  &\sim & e^{- i \o_- t -i l \f} (\cosh \r)^{-2 h_+} (\tanh \r)^l \,_2F_1 \left( h_+ + {l - \o_- \over 2}, h_+ + {l + \o_- \over 2}, l+1,\tanh^2 \r\right)\nonu
\tilde \F_{2m+1, 2n+1} &=& e^{-i \left( (n+{3 \over 4}) x_+ + (m+{3/4})x_-\right)} {\rm tr} \left( e^{-i\r  (L_1 + L_{-1})} \ket{2m+1}\bra{2n+1} \right)\nonu
 &\sim & e^{- i \o_+ t -i l \f} (\cosh \r)^{-2 h_+} (\tanh \r)^l \,_2F_1 \left(  h_+ + {l - \o_+ \over 2}, h_+ + {l + \o_+ \over 2}, l+1,\tanh^2 \r\right)\nonu
 \eea
 where $h_+ = 3/4, h_-=1/4$, and  we have defined
 \bea
 l &=& n-m\\
 \o_\pm &=& l + 2 h_\pm + 2m.
 \eea
The first identity in the above formulas follows  using the cyclicity of $\tr$, while in second identity we  evaluated the required traces using (\ref{traces}).
Comparing to  \cite{Balasubramanian:1998sn}, we see that $ \F_{2m, 2n} $ and $  \F_{2m+1, 2n+1}$  are precisely the positive frequency modes of the Klein-Gordon field in the `alternate' and `standard' quantizations respectively, with the correct frequency spectrum to guarantee  regularity in global $AdS_3$. The solutions
\be
\tilde \F_{0,0} = e^{-it /2} (\cosh\r)^{-1/2}, \qquad \tilde \F_{1,1} = e^{-3it /2} (\cosh\r)^{-3/2}
\ee
represent the lowest weight vectors in the decomposition (\ref{decomphshalf}), and
 the action of $hs[1/2 ]$  is realized on the wavefunctions  in terms of differential operators. For example, one can show that the generators of the $SL(2,
\RR) \times \widetilde{SL(2,\RR)}$ subgroup act  by  Lie derivatives \cite{Perlmutter:2012ds}.
One can perform a similar analysis for the mode solutions of the field $C$ leading  the  negative frequency solutions of the Klein-Gordon field in global $AdS_3$.
In the standard approach to AdS/CFT, one imposes  boundary conditions which  select only one of the representations in (\ref{decomphshalf}).

  It would be interesting to extend this analysis to the theories based on the higher spin algebra $hs[\l ]$ which have a realization in terms of deformed oscillators \cite{Vasiliev:1997dq}. When $\l$  equals $N$, a natural number greater than one, the generators $V^s_m$ with $s>N$ form an ideal which can be quotiented out, after which the higher spin algebra becomes $sl(N,\RR)$ and the associative algebra becomes the algebra of $N\times N$ matrices. In this case the single particle spectrum was analyzed in \cite{Perlmutter:2012ds}.

\section{Matter coupled to the higher spin square}\label{Sechss}
Now we turn to the example of interest, where the higher spin Lie algebra $\mh$ is the higher spin square $hss$ and  $\ma$ is the
underlying associative algebra $\ma_{hss}$, both of which we will presently describe following \cite{Gaberdiel:2015wpo}.
The algebra $\ma_{hss}$ has an oscillator realization in terms of the chiral modes\footnote{As in \cite{Gaberdiel:2015wpo}, we will ignore the zero modes and work in the sector of zero momentum and winding.} of the $\caln=4$ sigma model on $T^4$. This free superconformal field theory consists of two complex scalars and two complex  fermions with NS boundary conditions, leading to chiral oscillator modes\footnote{Our conventions are related to those of \cite{Ademollo:1976wv} as follows: starting from  the oscillators of  four real bosons, $b^i_m$, and four real fermions, $\s^i_m$, with $i = 1,\ldots ,4$, satisfying
$$ \left[ b^i_m, \left(b^j_n\right)^\dagger \right]= \d^{ij}\d_{mn}1, \qquad  \left\{\s^i_r, \left(\s^j_s\right)^\dagger \right\} = \d^{ij}\d_{rs}1$$
we have defined
\begin{align}
a^1_m &= {1\over \sqrt{2}} (b^1_m + i b^2_m), \qquad &  \bar a^1_m &= {1\over \sqrt{2}} (b^1_m - i b^2_m), \qquad &
a^2_m & = {1\over \sqrt{2}} (b^4_m + i b^3_m), \qquad & \bar a^2_m &= {1\over \sqrt{2}} (b^4_m - i b^3_m)\nonu
\psi^1_r &= {1\over \sqrt{2}} (\s^1_r + i \s^2_r), \qquad &  \bar \psi^1_r &= {1\over \sqrt{2}} (\s^1_r - i \s^2_r)\qquad &
\psi^2_r &= {1\over \sqrt{2}} (\s^4_r + i \s^3_r), \qquad & \bar a^2_r &= {1\over \sqrt{2}} (\s^4_r - i \s^3_r)\nonumber
\end{align}}
 $a^\a_m , \bar a^\a_m, \psi^\a_m, \bar \psi^\a_m$, with $m \in \NN_0, r\in \NN+\half, \a\in \{1,2 \}$, and their Hermitian conjugates. The canonical (anti-) commutation relations are
\begin{align}
 \left[ a^\a_m, \left(a^\b_n\right)^\dagger \right]&= \d^{\a\b}\d_{mn}1, &
 \left[\bar a^\a_m, \left(\bar a^\b_n\right)^\dagger \right]&= \d^{\a\b}\d_{mn}1 \\
\left\{\psi^\a_r, \left(\psi^\b_s\right)^\dagger \right\} &= \d^{\a\b}\d_{rs}1,&
\left\{\bar \psi^\a_r, \left(\bar \psi^\b_s\right)^\dagger \right\}&= \d^{\a\b}\d_{rs} 1\label{hssmodes}
\end{align}
with all other (anti-) commutators vanishing. The algebra $\ma_{hss}$ is the subalgebra of the oscillator algebra consisting of operators which annihilate both the in- and out Fock vacuum. A basis for $\ma_{hss}$ is formed
by the normal-ordered monomials in the oscillators which contain at least one creation and one annihilation operator. The algebra $hss$ is the Lie algebra obtained from $\ma$ by defining the Lie bracket to be  the commutator in $\ma$.

The algebra $hss$ contains a subalgebra $su(1,1|2)$, which arises as  the  vacuum-preserving subalgebra of the (small) $\caln=4$ superconformal algebra
which governs the $T^4$ theory  \cite{Ademollo:1976wv}.  The $su(1,1|2)$ algebra consists of $sl(2)$ generators $L_0,L_{\pm 1}$, $su(2)$ R-symmetry generators $J^i_0$ and superconformal generators $G^\a_{\pm \half}, \bar G^\a_{\pm \half}$. These are realized in terms of quadratic monomials in the oscillator  modes. The expressions for the conformal weight $L_0$
and R-charge generator $J^3_0$ are:
\bea
L_0 &=& \sum_{n=1}^\infty \left[ n\left(\left(a^\a_n\right)^\dagger a^\a_n  + \left(\bar a^\a_n\right)^\dagger \bar a^\a_n\right)+  \left(n -\half\right) \left( \left(\psi^\a_{n-\half}\right)^\dagger \psi^\a_{n-\half} +
\left(\bar \psi^\a_{n-\half}\right)^\dagger \bar \psi^\a_{n-\half} \right) \right]\nonu
J^3_0 &=& -\half \sum_{n=1}^\infty \left( \left(\psi^\a_{n-\half}\right)^\dagger \psi^\a_{n-\half} -
\left(\bar \psi^\a_{n-\half}\right)^\dagger \bar \psi^\a_{n-\half} \right),
\eea
where a sum over the index $\a =1,2$ is implied.  Expressions for the other generators can  be found in  \cite{Ademollo:1976wv}.

As in the previous example we would like to work out the decomposition (\ref{decomp}) of $\ma_{hss}$ into irreducible  $HSS \times \widetilde{HSS}$ representations, where the two copies
act from the right and left  respectively. The basis of $\ma_{hss}$ consisting of normal-ordered monomials is not
adapted to such a decomposition, since for example $L_0$ and $J^3_0$ don't act diagonally either from  the left or the right. Once again, a suitable basis for this purpose is the Fock  basis,
where we expand  general operators in basis elements of the form
\be \ket{F_1} \bra{F_2}, \ee
where $\ket{F_1} $ and $\ket{F_2}$ are  elements of the Fock space $\calf$ built from acting with the creation modes in (\ref{hssmodes}) on the vacuum.

The  transformation  between the normal ordered monomial and Fock  bases can be worked out explicitly.
For notational simplicity, let us illustrate this for a single bosonic oscillator mode $a$ and for a  fermionic mode $\psi$. 
For a bosonic oscillator, the relation between normal-ordered and Fock-ordered monomials is summarized by (see Appendix \ref{App} for more details):
\bea
[ (a^\dagger)^m a^n]_F &=& : (a^\dagger)^m a^n e^{- a^\dagger a} :\nonu
(a^\dagger)^m a^n &=& [(a^\dagger)^m a^n e^{ a^\dagger a}]_F. \label{FockNormalbos}
\eea
For a fermionic oscillator, similar formulas hold:
\bea
\, [ (\psi^\dagger)^m \psi^n]_F &=& : (\psi^\dagger)^m \psi^n e^{- \psi^\dagger \psi} :\nonu
\, (\psi^\dagger)^m \psi^n &=& [(\psi^\dagger)^m \psi^n e^{ \psi^\dagger \psi}]_F,\label{FockNormalferm}
\eea
where  most of the terms in the expansion of the right hand side actually vanish due to $(\psi^\dagger)^2 = \psi^2=0$.
Returning to the full algebra $\ma_{hss}$, the monomial and Fock operator bases can be worked out in principle by repeated application of  (\ref{FockNormalbos},\ref{FockNormalferm}). Note that in the Fock  basis, both $L_0$ and
$J^3_0$ act diagonally both from the left and right.

We have not yet imposed, on our Fock operator basis,  the restriction that that
elements of $\ma_{hss}$  should annihilate the in- and out- Fock vacua. Decomposing the Fock space $\calf$ as
\be
\calf = \CC \ket{0} \oplus \calf',\label{decomprep}
\ee
where $\calf'$ is the direct sum of the one-, two-, and more-particle Hilbert spaces,  a basis for $\ma_{hss}$  is formed by elements of the form
\be
\ket{F_1'} \bra{F_2'} \qquad {\rm with \ } \ket{F_1'}, \ket{F_2' } \in \calf' .\label{basishss}
\ee
It follows that, as a vector space, $\ma_{hss}$ is simply
\be
\ma_{hss} = \calf' \otimes {\calf'}^* \label{decompfock}
\ee
Viewed as representations of $hss$, the $\CC \ket{0}$ term in the decomposition (\ref{decomprep}) corresponds to the the trivial representation, while  the second term $\calf'$  carries a representation
which turns out to be the minimal representation $V_{min}$ discussed in \cite{Gaberdiel:2014cha},\cite{Gaberdiel:2015wpo}.
Indeed, from the above considerations we easily compute the character
\bea
\chi_{\calf '}&\equiv&  \tr_{\calf '}  q^{L_0}y^{2 J^3_0}\nonu
&=& \tr_{\calf }  q^{L_0}y^{2 J^3_0} -1\nonu
&=& Z_{T^4}-1
\eea
where $Z_{T^4}$ is the chiral partition function (\ref{ZT4}) of the  $T^4$ SCFT. This  agrees with the character of the  minimal representation (\ref{chimin}) of $hss$  \cite{Gaberdiel:2014cha},\cite{Gaberdiel:2015wpo}.

Combining (\ref{decompfock}) with the observation that $\calf ' = V_{min}$  we arrive at the following simple decomposition (\ref{decomp}) in the case of the higher spin square:
\be
\ma_{hss} = ( V_{min}, V_{min}^* ).\label{decomphss}
\ee
The positive frequency solutions which furnish this representation are, from (\ref{gensolAdS}),
\be
\tilde C_{F_1', F_2'} = e^{-i( h_{F_1'} x_- + h_{F_2'} x_+ )} \tilde R \ket{F_1'} \bra{F_2'} R^{-1}
\ee
where $h_F$ denotes the $L_0$-eigenvalue of $\ket{F}$.
From (\ref{1partspectr}) we read off that the partition function of the single particle states  of the matter field is
\be
Z^{1-part}_{hss} (q,\bar q, y, \bar y)  = \tr_{\ma_{hss}} q^{L_0} \bar q^{\tilde L_0} y^{2J^3_0}\bar y^{2\tilde J^3_0}= |\chi_{min}(q,y)|^2.
\ee
which reproduces (\ref{Z1part}).
For completeness, let us also indicate how the multi-particle spectrum, keeping track of statistics, leads to the full CFT result (\ref{Zmatter}).
We define degeneracies $c(h, \tilde h, l, \tilde l)$ from writing the single-particle partition function as
\be
Z^{1-part}_{hss}  = \sum_{h,\tilde h, l, \tilde l} c(h, \tilde h, l, \tilde l) q^h \bar q^{\tilde h} y^l \bar y^{\tilde l}.
\ee
We further note that $c(h, \tilde h, l, \tilde l)$ counts bosonic states when $l+\tilde l$ is even and fermions when  $l+\tilde l$ is odd. Therefore the multiparticle partition function can be written as
\be
Z_{hss}^{matter} = \prod_{h, \tilde h} {\prod_{l,\tilde l; l+\tilde l \ {\rm odd}} (1+ q^h \bar q^{\tilde h} y^l \bar y^{\tilde l})^{c(h, \tilde h, l, \tilde l)} \over \prod_{l,\tilde l; l+\tilde l \ {\rm even}} (1- q^h \bar q^{\tilde h} y^l \bar y^{\tilde l})^{c(h, \tilde h, l, \tilde l)}}.
\ee
After  some algebra this can be rewritten in the form (\ref{Zmatter}).

We note that, unlike in the $hs[1/2]$ case (\ref{decomphshalf}), the decomposition (\ref{decomphss}) contains only one term, so that unlike in that example we don't have a choice of several  boundary conditions which are compatible with the $hss$ symmetry.
This seems similar to what happens in other  examples (see e.g. \cite{Klebanov:1999tb}) where only one of the two possible boundary conditions on a scalar is compatible with supersymmetry.

\section{Generalization and outlook}
In this note we have confirmed Gaberdiel and Gopakumar's conjecture that the full contribution (\ref{Zmatter}) to the untwisted  sector CFT partition function arises  from a single matter field in the bulk, which is furthermore  coupled to the $hss$  gauge fields in a straightforward generalization of Vasiliev's linearized theory \cite{Vasiliev:1992gr}.
It would be interesting to understand better how this spectrum decomposes in terms of supermultiplets of the AdS supergroup $SU(1,1|2) \times \widetilde{SU(1,1|2)}$ (see \cite{Gunaydin:1986fe}) contained in $HSS \times \widetilde{HSS}$.
It would also be of interest to understand how, after imposing appropriate boundary conditions on the gauge fields as in \cite{Henneaux:2010xg},\cite{Campoleoni:2010zq}, the spectrum of boundary excitations reproduces the extended vacuum character (\ref{Zgauge}).

The symmetric orbifold CFT also contains  twisted sectors, whose decomposition into $HSS \times \widetilde{HSS}$   representations is not fully understood at present.  We note that any representation $(R, \tilde R)$ entering
in such a decomposition  can be described in the bulk by fields $C$ and $\tilde C$ taking values in $R \otimes \tilde R$ and $\tilde R \otimes R$ respectively, with linearized equations
\bea
d  C -  A_R  C + C \tilde  A_{\tilde R} &=&0\nonu
d  \tilde C - \tilde A_{\tilde R} \tilde  C + \tilde  C   A_R &=&0
\eea
where $A_R $ and  $\tilde A_{\tilde R}$ are the gauge fields in the representations $R, \tilde R$ respectively.

These considerations show how the full spectrum of the symmetric orbifold can in principle be reproduced from linearized matter fields in the  bulk. The main open question is of course if and how these fields can be incorporated into a consistent  interacting theory which furthermore  reproduces the correlation functions of the symmetric orbifold. The hope is that the $hss$ symmetry will be sufficiently restrictive that to determine the interactions essentially uniquely, as is the case for the Vasiliev theories.

\section*{Acknowledgements}
 I would like to  thank  R. Gopakumar, C. Iazeolla, E. Perlmutter and T. Proch\'azka for helpful comments and discussions.  This research was supported by the Grant Agency of the Czech Republic under the grant 14-31689S.

\bigskip

\begin{appendix}

\section{Some ordering  identities}\label{App}
In this section we will work out some relations between different ordering prescriptions for operators constructed out of a single harmonic oscillator needed in the main text.  In particular, we need to relate the Fock  basis to
the basis of Weyl-ordered monomials in section \ref{Sechshalf}  and to the basis of normal ordered monomials  in section \ref{Sechss}.
It's useful to define a  `Fock ordering'  as the operation which   turns  monomials in the variables $a, a^\dagger$  into operators by letting the creation and annihilation operators act on $\ket{0}  \bra{0} $ from the left and  the right respectively, e.g.
\be
[(a^\dagger)^m a^n]_F = (a^\dagger)^m \ket{0}  \bra{0} a^n.
\ee
Any operator can be expanded in the bases of Fock-, normal-  or  Weyl-ordered monomials as
\be
\calo = \left[f^F_\calo (a, a^\dagger)\right]_F= :f^N_\calo (a, a^\dagger): = \left[f^W_\calo (a, a^\dagger)\right]_W
\ee
where $f^F_\calo, f^N_\calo, f^W_\calo$ are  the  Fock-, normal-  and  Weyl-ordered symbols respectively. These are given by
\bea
 f^F_\calo (z, \bar z) &=& \sum_{m,n\in \NN} {\bra{m} \calo \ket{n}\over \sqrt{m! n!}}  \bar z^m z^n \label{Focksymbol}\\
 f^N_\calo (z, \bar z) &=&\bra{z} \calo \ket{z} \label{normalsymbol}\\
 f^W_\calo (z, \bar z) &=& {2 e^{2 |z|^2}\over \p} \int d^2 w \bra{-w}\calo \ket{w} e^{-2( w\bar z- \bar w z)} \label{Weylsymbol}
 \eea
 where we have made use of the  coherent states
 \be
 \ket{z} = e^{-\half z \bar z} e^{z a^\dagger} \ket{0}.
 \ee
 For a derivation of  last two expressions (\ref{normalsymbol},\ref{Weylsymbol})  see e.g. \cite{Agarwal:1971wc}.

 From (\ref{Focksymbol},\ref{normalsymbol}) we find the following simple relation between the normal- and Fock-ordered symbols:
 \bea
 f^F_\calo (z, \bar z) &=& f^N_\calo (z, \bar z) e^{|z|^2}\nonu
  f^N_\calo (z, \bar z) &=& f^F_\calo (z, \bar z) e^{-|z|^2}.\label{Focknormalsymbol}
 \eea
 Applied to monomial basis elements these lead to the expressions (\ref{FockNormalbos}).

 Next we wish to derive a relation between Weyl-ordered and Fock-ordered symbols. We start from  the expressions relating Weyl- and normal-ordered symbols\footnote{These formal expressions are to be treated with caution as convergence issues may arise, see \cite{Widder}
 for more details.}
 (see (5.15) and (5.30) in \cite{Agarwal:1971wc})
 \bea
 f^N_\calo (z, \bar z) &=&  {2\over \p} \int d^w f^W_\calo (w, \bar w)  e^{-2 |w-z|^2}\nonu
 f^W_\calo (z, \bar z) &=& e^{- \half \pa_z \pa_{\bar z} } f^N_\calo (z, \bar z)
 \eea
 Combining  these with (\ref{Focknormalsymbol}) we  obtain a relation between Weyl-ordered and Fock-ordered symbols:
 \bea
f^F_\calo (z, \bar z) &=& {2 e^{|z|^2} \over \p} \int d^2 w f^W_\calo (w, \bar w) e^{-2 |w-z|^2}\nonu
f^W_\calo (z, \bar z) &=& e^{- \half \pa_z \pa_{\bar z}} \left( f^F_\calo (z, \bar z) e^{- |z|^2}\right).
\eea
Application of the first equation leads, after some algebra, to the expansion of Weyl-ordered basis elements in terms of Fock-ordered ones (\ref{WeyltoFock}). For the inverse relation, instead of using the second equation it is more convenient to  apply (\ref{Weylsymbol}), leading to (\ref{FocktoWeyl}).

From (\ref{Weylsymbol}) we also derive an expression \cite{Kraus:2012uf} for the trace operation, defined as extracting the coefficient of the unit operator in an expansion in Weyl-ordered monomials,
\be
\tr \calo = {2 \over \p} \int d^2 z \bra{ -z } \calo \ket{z}  .
\ee
From this we compute the  following traces needed in section \ref{Sechshalf}:
\begin{align}
&{\rm tr} \left( e^{-i\r  (L_1 + L_{-1})} \ket{m}\bra{n} \right) =&\nonu
& \caln_{m,n} (\cosh \r)^{-{3 \over 2}} (\tanh \r)^{{n-m\over 2}} \,_2 F_1
\left({1-m\over 2}, {n \over 2} +1,{n-m \over 2} + 1, \tanh^2 \r\right)&\label{traces}
\end{align}
where $\caln_{m,n}$ is a constant which is irrelevant for our purposes.
To derive this result we made  use of the BCH-type rearrangement formula
\be
e^{i\m  (L_1 + L_{-1})} = e^{ i \tanh \m L_{-1}} e^{ i \cosh \m \sinh \m L_{1}} (\cosh \m)^{-2 L_0}.
\ee
\end{appendix}

\end{document}